\documentclass[aps,twocolumn,prb,showpacs]{revtex4}
\usepackage{epsfig}
\usepackage{bm}
\def\be{\begin{equation}}
\def\ee{\end{equation}}
\def\bea{\begin{eqnarray}}
\def\eea{\end{eqnarray}}
\def\nn{\nonumber}
\def\upa{\uparrow}
\def\dwa{\downarrow}

\begin{document}
\title{Correlation Amplitudes for the spin-$1/2$ $XXZ$ chain 
in a magnetic field}
\author{T. Hikihara}
\author{A. Furusaki}
\affiliation{Condensed-Matter Theory Laboratory, RIKEN,
 Wako, Saitama 351-0198, Japan}
\date{\today}

\begin{abstract}
We present accurate numerical estimates for the correlation
amplitudes of leading and main subleading terms of the
two- and four-spin correlation functions in the one-dimensional
spin-$1/2$ $XXZ$ model under a magnetic field.
These data are obtained by fitting the correlation functions,
computed numerically with the
density-matrix renormalization-group method,
to the corresponding correlation functions in the low-energy
effective theory.
For this purpose we have developed the Abelian bosonization
approach to the spin chain under the open boundary conditions.
We use the numerical data of the correlation amplitudes
to quantitatively estimate spin gaps induced by a transverse
staggered field and by exchange anisotropy.
\end{abstract}

\maketitle

\section{Introduction}

The one-dimensional spin-$1/2$ $XXZ$ model is a basic and
well studied model in statistical physics.
In some parameter range its ground state is critical and
spin-spin correlation functions exhibit quasi-long-range order.
The model represents a typical example of 
Tomonaga-Luttinger (TL) liquids in which elementary excitations 
are gapless collective modes rather than single-particle excitations,
in contrast to three-dimensional systems.
These features are uncovered by extensive theoretical studies
over the years which have employed various
powerful methods including the Bethe ansatz,\cite{Baxter} 
bosonization technique,\cite{GogolinNT}
and the conformal field theory.\cite{Affleck}
One-dimensional spin chains are also important from the experimental
viewpoint, as they are relevant to many quasi-one-dimensional magnets
in which couplings along one direction are considerably stronger 
than those in the other two directions.

In this paper, we discuss long-distance asymptotes of equal-time
correlation functions in the spin-$1/2$ $XXZ$ chain in a magnetic field.
The Hamiltonian is
\be
\mathcal{H}_0 = J \sum_{l}
            \left( S^x_l S^x_{l+1} + S^y_l S^y_{l+1}
                   + \Delta S^z_l S^z_{l+1} \right)
- H \sum_{l} S^z_l,
\label{eq:Ham}
\ee
where $J>0$ and
${\bm S}_l=(S_l^x,S_l^y,S_l^z)$ is an $S=1/2$ spin operator
on the $l$th site of the chain.
The anisotropy parameter $\Delta$ is assumed to satisfy the inequality
 $\Delta > -1$ 
such that the system is in the critical regime 
for a certain range of magnetic fields, 
i.e., $0 \le |H| < H_{c2}$ for $-1 < \Delta \le 1$ 
and $H_{c1} < |H| < H_{c2}$ for $\Delta > 1$ where 
$H_{c1}$ and $H_{c2}$ are the lower and upper critical fields,
respectively.
The spins partially polarized by the field $H$ have a finite
magnetization per site $M$, where $-1/2<M<1/2$.
The low-energy excitations of
$\mathcal{H}_0$ in the critical regime are free massless bosons.
They are governed by the Gaussian theory
\be
\widetilde{\mathcal{H}}_0 = \frac{v}{2} \int dx
\left[ \left(\frac{d\phi}{dx}\right)^2
+ \left(\frac{d\tilde{\phi}}{dx}\right)^2 \right],
\label{eq:Gauss}
\ee
where $v$ is the renormalized spin-wave velocity and 
the bosonic fields $\phi(x)$ and $\tilde{\phi}(x)$ obey
the commutation relation 
$[ \phi(x), \tilde{\phi}(y)] = -(i/2) [1 + {\rm sgn}(x-y)]$.
We take the lattice spacing $a=1$ and identify the site index $l$
with the continuous space variable $x$.
For a better description of low-energy physics, one needs to add
to $\widetilde{\mathcal{H}}_0$ 
an irrelevant operator $\cos(2\phi/R)$: 
This operator becomes marginal at $\Delta =1$ and $H = 0$ yielding 
logarithmic corrections, while it is relevant and opens a spin gap 
for $\Delta > 1$ and $0 < |H| < H_{c1}$.
We ignore this term in this paper, however.
This may introduce systematic errors in our analysis 
for $\Delta \agt 1$ and small $M$.
The spin operators are related to the bosonic
fields\cite{Affleck,SheltonNT,EggertA} as 
\bea
S^z_l &=& M + \frac{1}{2\pi R} \frac{d\phi}{dx}
\nn \\
&&~~~- a_1 (-1)^{l} 
  \sin\left(Q l + \frac{\phi(l)}{R}\right) + \cdots,  
\label{eq:Szintro} \\
S^+_l &=& e^{ i 2\pi R \tilde{\phi}(l)} 
\!\left[
b_0 (-1)^l + b_1 \sin\left(Q l + \frac{\phi(l)}{R} \right) + \cdots
\right],
\nn \\
\label{eq:Sxintro}
\eea
where
$Q = 2 \pi M$ is incommensurate wave number,
and $a_n$ and $b_n$ are non-universal constants which depend on
$\Delta$ and $M$.
The TL liquid parameter (or the compactification radius) $R$
characterizes the asymptotic behavior of 
correlation functions and its exact value is readily
obtained by solving the Bethe ansatz integral
equations.\cite{BogoIK,QinFYOA,CabraHP}
Essentially all the low-energy properties of the spin chain
(\ref{eq:Ham}) 
follow from Eqs.\ (\ref{eq:Gauss})--(\ref{eq:Sxintro}).
For example, the ground-state two-spin correlation functions 
are found to decay algebraically,\cite{LutherP,ladder}
\bea
\langle S_l^x S_{l'}^x \rangle &=&
A^x_0 \frac{(-1)^{l-l'}}{|l-l'|^\eta}
-A^x_1 \frac{\cos[Q(l-l')]}{|l-l'|^{\eta+1/\eta}} + \cdots,
\label{eq:tdlcx} \\
\langle S_l^z S_{l'}^z \rangle &=&
M^2 -\frac{1}{4\pi^2\eta|l-l'|^2}
\nn \\
& &~~~+ A^z_1 (-1)^{l-l'}\frac{\cos[Q(l-l')]}{|l-l'|^{1/\eta}} + \cdots.
\label{eq:tdlcz}
\eea
The decay exponent $\eta = 2\pi R^2$ is a function of $\Delta$ and $M$.
For $-1 < \Delta \le 1$, $\eta = 1-\cos^{-1}(\Delta) /\pi$ at $M=0$,
while $\eta = 2$ at $M\to+0$ for $\Delta > 1$.
As $M$ increases, $\eta$ varies monotonically and approaches
the universal value $1/2$ in the limit $M \to 1/2$.
The amplitudes $A^x_n$ and $A^z_n$ are related to
the coefficients $a_n$ and $b_n$.
Recently Lukyanov and coworkers\cite{LukyanovZ,Lukyanov,LukyanovT} have
obtained exact formulas of the correlation amplitudes
in Eqs.\ (\ref{eq:tdlcx}) and (\ref{eq:tdlcz}) at $M=0$ 
and $-1 < \Delta \le 1$.
Nevertheless their values at $M\ne0$ are not known analytically.

The aim of this paper is to numerically determine
the non-universal coefficients $a_n$ and $b_n$ with high accuracy
for arbitrary $M$.
We first extend the bosonic representation of 
the spin operators (\ref{eq:Szintro}) and (\ref{eq:Sxintro})
to a more general form of infinite series, and calculate
the spin polarization  $\langle S^z_l \rangle$ and
the two-spin correlation functions 
$\langle S^x_l S^x_{l'} \rangle$ and $\langle S^z_l S^z_{l'} \rangle$
within the bosonization theory.
The analytic formulas so obtained are used to fit numerical data
which we compute by using the density-matrix 
renormalization group (DMRG) method.\cite{White1,White2} 
The fitting parameters are the coefficients $a_n$ and $b_n$.
This scheme is basically the same as the one used in our previous 
studies,\cite{corAmp,ladder} and in this paper we provide
more accurate numerical data of the coefficients $a_1$, $b_0$, and
$b_1$ for wider range of parameters.

The information on the coefficients is crucial for
quantitative analysis on effects of perturbations
to which the critical ground state of the spin-1/2 $XXZ$ model
has instabilities.
Such perturbations include bond alternation,\cite{bond1,bond2} 
next-nearest-neighbor coupling,\cite{NNN1,NNN2,NNN3} 
and transverse staggered magnetic 
field.\cite{OshikawaA,AffleckO,EsslerT,Essler}
In the bosonization approach, a perturbation written in the original
spin operators is translated into bosonic operators
through Eqs.\ (\ref{eq:Szintro}) and (\ref{eq:Sxintro}).
The impact of the perturbation on the ground state,
the vacuum state of the Gaussian model (\ref{eq:Gauss}),
can be qualitatively estimated from scaling
dimension of the perturbation operators,
which depends only on the TL liquid parameter $R$.
It is, however, necessary to know the exact values of
the non-universal coefficients $a_n$ and $b_n$ as well, to quantitatively
analyze effects of the perturbation,
e.g., to estimate magnitude of an energy gap induced by the perturbation
or to compute correlation functions in the presence of the perturbation.
As an example of such quantitative analysis we calculate the spin gap
induced by the staggered transverse 
field\cite{OshikawaA,AffleckO,EsslerT,Essler}
from the numerical data of the coefficients.

There are often cases when perturbations are composite operators
of two (or more) spins, such as exchange interactions.
In this case one must consider fusion of two operators.
In general, fusing two spin operators at short distances 
may generate operators 
which are not present in each single-spin operator which are fused.
Consequently, the leading term of the correlator of the composite
operator may be different from
a product of the leading terms of the correlators of
the single-spin operators.
We illustrate this by taking the nearest-neighbor coupling 
$S^x_lS^x_{l+1}$ as an example.
We examine operators generated by the fusion of
$S^x_l$ and $S^x_{l+1}$.
From a numerical fitting of four-spin correlation functions,
we estimate the amplitude of a leading uniform term of 
the correlation function.
The results are used to analyze the spin gap 
induced by the perturbation of exchange anisotropy.\cite{DmitrievKO,CauxEL}

The paper is organized as follows.
We briefly review the Abelian bosonization approach to 
the model (\ref{eq:Ham}) in the next section.
The bosonic representation of spin operators in the form of infinite
series is introduced in Sec.\ II A.
We then use it to derive the analytic formulas of the two- and 
four-spin correlation functions in Sec.\ II B and C.
In Sec.\ III, we present the numerical results on the correlation
functions obtained from the DMRG calculation.
The numerical data of amplitudes of the two- and four-spin correlation
functions are presented in Sec.\ III A and B, respectively.
We show that the numerical results are in excellent agreement
with analytical predictions available for various limiting cases.
Furthermore, we discuss the spin gaps 
induced by the perturbations of staggered transverse field and 
exchange anisotropy in Sec. III C.
Finally, we summarize the results in Sec.\ IV.
Appendix explains bosonization of spin operators.

\section{Bosonization}

\subsection{Spin operators}

In this section, we summarize the Abelian bosonization technique 
applied to the spin-1/2 $XXZ$ chain in a magnetic field.
Let us first express the original spin operators 
$S^x_l$ and $S^z_l$ in terms of the bosonic fields.
To this end, we follow and extend the scheme
of Refs.~\onlinecite{Affleck} and \onlinecite{SheltonNT}.
We begin with the bosonic representation of electron operators 
in the Hubbard model with the on-site repulsion $U$.
At half filling we have a gapped charge mode and a gapless spin mode.
After constructing spin operators from the electron operators, 
we integrate out the gapped charge mode
to obtain a bosonic representation of the spin operators 
in the Heisenberg chain.
We then generalize the result to the $XXZ$ case.
A detailed derivation is presented in Appendix, 
and here we show only the final results, 
\bea
S^z_l &=&
\frac{1}{2\pi R}\frac{d\phi}{dx}
-\sum_{n=0}^\infty a_{2n+1} (-1)^l
 \sin\!\left[(2n+1)\frac{\phi(x)}{R}\right],
\quad
\label{eq:Sz}  \\
S^+_l &=& e^{i2\pi R \tilde{\phi}(x)} 
         \sum_{n=0}^\infty \left\{
           b_{2n} (-1)^l \cos\!\left[ 2n 
              \frac{\phi(x)}{R}\right]\right.
\nn \\
&& + \left. b_{2n+1} \sin\!\left[ (2n+1) 
              \frac{\phi(x)}{R}\right]
\right\},
\label{eq:S+}
\eea
where $a_n$ and $b_n$ are non-universal constants.
They depend on the short-distance regularization of the Gaussian theory
as well as on the parameters $\Delta$ and $M$.
Equations (\ref{eq:Szintro}) and (\ref{eq:Sxintro}) are just the first
few terms of Eqs.\ (\ref{eq:Sz}) and (\ref{eq:S+})
with the shift of the field $\phi(x)\to\phi(x)+QRx$.
In principle the right hand side of Eqs.\ (\ref{eq:Sz}) and (\ref{eq:S+})
should contain descendant fields as well.
It follows from Eq.\ (\ref{eq:S+}) that
\bea
S^x_l &=& \frac{1}{2}(S^+_l+S^-_l)=
\frac12 \left[S^+_l+(S^+_l)^\dagger\right]
\nn \\
 &=& \sum_{n=0}^\infty 
\left\{ b_{2n} (-1)^l \cos\!\left[2 \pi R \tilde{\phi}(x)\right]
\cos\!\left[ 2n \frac{\phi(x)}{R} \right] \right. 
\nn \\
& &\left.~~~~+ i b_{2n+1} \sin\!\left[2 \pi R \tilde{\phi}(x)\right]
\sin\!\left[ (2n+1) \frac{\phi(x)}{R} \right] \right\}.
\nn \\&&
\label{eq:Sx}
\eea
Here we have used the commutator $[\phi(x),\tilde\phi(x)]=-i/2$.
An equivalent bosonic representation of the spin operators is
recently derived in Ref.\ \onlinecite{LukyanovT} from 
global symmetry analysis of the lattice and field operators.
Our derivation in Appendix is complementary
to Ref.\ \onlinecite{LukyanovT}.

\subsection{Two-spin correlation functions}

The two-spin correlation functions are readily
calculated by the bosonization method.
Let us first consider the thermodynamic limit where the correlation
functions depend only on the distance between the two spins.
In this case we expand the bosonic fields 
$\phi(x)$ and $\tilde{\phi}(x)$ as 
\bea
\phi(x) &=&
\int_0^\infty\! dk \frac{e^{-\lambda k/2}}{\sqrt{4\pi}}
\left[ \frac{e^{ ikx}}{\sqrt{k-i\delta}} (\alpha_k + \alpha^\dagger_{-k}) 
\right.
\nn \\
&&\left.~~~~~~~~~~
     + \frac{e^{-ikx}}{\sqrt{k+i\delta}} (\alpha^\dagger_k + \alpha_{-k})
\right]
+QRx,\quad
\label{eq:FTphi} \\
\tilde{\phi}(x) &=&
\int_0^\infty\! dk \frac{e^{-\lambda k/2}}{\sqrt{4\pi}}
\left[ \frac{e^{ ikx}}{\sqrt{k+i\delta}} (-\alpha_k + \alpha^\dagger_{-k}) 
\right.
\nn \\
&&\left.~~~~~~~~~~~~~~~~
     + \frac{e^{-ikx}}{\sqrt{k-i\delta}} (-\alpha^\dagger_k + \alpha_{-k})
\right],
\label{eq:FTtphi}
\eea
where $\alpha_k$ and $\alpha^\dagger_k$ are boson operators 
obeying the commutation relation 
$[ \alpha_k, \alpha^\dagger_{k'}] = \delta(k-k')$,
$\lambda$ is a short-distance cutoff,
and $\delta$ is a positive infinitesimal.
The fields $\phi(x)$ and $\tilde\phi(x)$ defined by Eqs.\ (\ref{eq:FTphi})
and (\ref{eq:FTtphi}) satisfy the commutation relation
$[\phi(x),\tilde\phi(y)]=-i\Theta(x-y)$ in the limit $\lambda\to0$,
where $\Theta(x)$ is the step function.
The Gaussian Hamiltonian (\ref{eq:Gauss}) defined on the whole real
$x$ axis now reads
\be
\widetilde{\mathcal{H}}_0 
= \int_0^\infty\!\! vk (\alpha^\dagger_k \alpha_k 
                  + \alpha^\dagger_{-k} \alpha_{-k}) dk
+ \mathrm{const.}
\ee
Thus, the ground state $|0 \rangle$ of the Hamiltonian (\ref{eq:Ham}) 
corresponds to the vacuum for the bosons $\alpha_k$,
i.e., $\alpha_k |0\rangle = 0$.
Substituting Eqs.\ (\ref{eq:FTphi}) and (\ref{eq:FTtphi}) to
Eqs.\ (\ref{eq:Sz}) and (\ref{eq:Sx}), 
we obtain the equal-time two-point correlators
in the thermodynamic limit,
\bea
\langle S^x_l S^x_{l'} \rangle &=&
\sum_{n=0}^\infty 
\left\{ 
A^x_{2n} (-1)^{l-l'} \frac{\cos[2nQ(l-l')]}{|l-l'|^{\eta+(2n)^2/\eta}} 
\right.
\nn \\
& &\left. ~~~
- A^x_{2n+1}\frac{\cos[(2n+1)Q(l-l')]}{|l-l'|^{\eta+(2n+1)^2/\eta}}
 \right\},
\label{eq:Cxtdl1} \\
\langle S^z_l S^z_{l'} \rangle &=& 
M^2 - \frac{1}{4\pi^2 \eta |l-l'|^2} 
\nn \\
&& 
+\sum_{n=0}^\infty 
A^z_{2n+1} (-1)^{l-l'} \frac{\cos[(2n+1)Q(l-l')]}{|l-l'|^{(2n+1)^2/\eta}},
\nn \\
\label{eq:Cztdl1}
\eea
where $\langle \cdots \rangle=\langle0|\cdots|0\rangle$
represents the expectation value in the lowest energy state 
in the sector with magnetization $M$.
If we adopt the regularization 
\be
\int^\infty_0\! dk\frac{e^{-\lambda k}}{k}(1-\cos kx)=\ln x,
\label{eq:regularization1}
\ee
then the correlation amplitudes are related to the coefficients 
in Eqs.\ (\ref{eq:Sz}) and (\ref{eq:Sx}) by
\be
A^x_n = \frac{b_n^2}{4}(1+\delta_{n,0}),
\quad
A^z_n = \frac{a_n^2}{2}.
\ee
For $M = 0$ and $-1 < \Delta < 1$, the values of the amplitudes 
$A^x_0$, $A^x_1$, and $A^z_1$ are obtained by
analytical\cite{McCoy,LukyanovZ,Lukyanov,LukyanovT} 
and numerical\cite{corAmp} methods.
We note that in principle the long-distance expansion (\ref{eq:Cxtdl1})
and (\ref{eq:Cztdl1}) should also include contributions from the
descendants ignored in Eqs.\ (\ref{eq:Sz}) and (\ref{eq:Sx}) and
those from the irrelevant operators discarded in
$\widetilde\mathcal{H}_0$.\cite{LukyanovT}

As mentioned in Introduction, we determine the coefficients $a_n$ and
$b_n$ by fitting numerically computed correlation functions to
appropriate formulas.
Since the numerical DMRG method works best for finite-size systems
with open boundaries, we need calculate the correlation functions
under the open boundary conditions.
For this purpose we employ the open-boundary bosonization scheme
developed in Refs.\ \onlinecite{corAmp} and \onlinecite{ladder}.
Suppose that the $XXZ$ spin chain of our interest consists of
$L$ spins $\bm{S}_l$ ($l=1,2,\ldots,L$).
This is equivalent to assume $\bm{S}_0=\bm{S}_{L+1}=0$.
In the bosonic representation this amounts to have the Gaussian model
(\ref{eq:Gauss}) defined in the finite region $0<x<L+1$ with
Dirichlet boundary conditions at $x=0$ and $x=L+1$.
Our convention is that
\be
\phi(0)=0, \quad \phi(L+1)=2\pi RLM.
\label{eq:bc}
\ee
These boundary conditions are consistent with
Eqs.\ (\ref{eq:Sz})--(\ref{eq:Sx}).
Once we fix the boundary
conditions (\ref{eq:bc}) and the regularization scheme,
the coefficients $a_n$ are uniquely determined whereas the coefficients
$b_n$ are determined only up to a phase factor.
Instead of Eqs.\ (\ref{eq:FTphi}) and (\ref{eq:FTtphi}), 
we now have the mode expansion\cite{EggertA,ladder,corAmp}
\bea
\phi(x) &=& \frac{x}{L+1}\phi_0
+ \sum_{n=1}^{\infty} \frac{\sin(q_nx)}{\sqrt{\pi n}} 
     (\tilde{\alpha}_n + \tilde{\alpha}_n^\dagger ),
\label{eq:phi} \\
\tilde{\phi}(x) &=& \tilde{\phi}_0
+ i \sum_{n=1}^{\infty} \frac{\cos(q_nx)}{\sqrt{\pi n}} 
     (\tilde{\alpha}_n - \tilde{\alpha}_n^\dagger ),
\label{eq:tphi}
\eea
where $q_n = \pi n/(L+1)$, 
$[ \tilde{\phi}_0, \phi_0 ] = i$, and 
$\tilde{\alpha}_n$ and $\tilde{\alpha}^\dagger_n$ are boson operators
satisfying $[\tilde\alpha_n,\tilde\alpha_{n'}^\dagger]=\delta_{n,n'}$.
The Gaussian model (\ref{eq:Gauss}) becomes
\be
\widetilde\mathcal{H}_0=
\sum^\infty_{n=1}vq_n\tilde\alpha^\dagger_n\tilde\alpha_n
+\frac{v\phi_0^2}{2(L+1)}-\frac{\pi v}{24(L+1)}.
\ee
The lowest-energy state of the spin chain (\ref{eq:Ham})
with magnetization $M$ corresponds to a vacuum of bosons
$\tilde{\alpha}_n |0 \rangle = 0$
with $\phi_0 |0 \rangle = 2\pi RLM|0\rangle$.
To calculate the zero-temperature correlation functions,
we substitute Eqs.\ (\ref{eq:phi}) and (\ref{eq:tphi})
into Eqs.\ (\ref{eq:Sz}) and (\ref{eq:Sx})
and take average with respect to the state $|0 \rangle$.
(The readers interested in the detailed calculation should refer 
to Ref.\ \onlinecite{ladder}.)
We obtain
\begin{widetext}
\bea
\langle S_l^x S_{l'}^x \rangle
&=& \frac{f_{\frac{\eta}{2}}(2l) f_{\frac{\eta}{2}}(2l')}
         {f_\eta(l-l') f_\eta(l+l')}
 \sum_{n,n'=0}^\infty 
 \frac{s(n,n';l,l') b_n b_{n'}}
      {4 f_{\frac{n^2}{2\eta}}(2l) f_{\frac{{n'}^2}{2\eta}}(2l')}
\left\{ 
 \cos\left[q(nl+n'l')\right] \frac{f_{\frac{nn'}{\eta}}(l-l')}
                       {f_{\frac{nn'}{\eta}}(l+l')}
+\cos\left[q(nl-n'l')\right] \frac{f_{\frac{nn'}{\eta}}(l+l')}
                       {f_{\frac{nn'}{\eta}}(l-l')}
\right\},
\nn \\
&&\label{eq:Cxfin} \\
\langle S^z_l S^z_{l'} \rangle &=&
    \left( \frac{q}{2\pi} \right)^2
     - {\sum_{n=1}^\infty}' \frac{qa_{n}}{2\pi}
     \left[ \frac{(-1)^l \sin\left(n q l\right)}
                 {f_{\frac{n^2}{2\eta}}(2l)}
   + \frac{(-1)^{l'} \sin\left(n q l'\right)}
          {f_{\frac{n^2}{2\eta}}(2l')} 
   \right]
- \frac{1}{4\pi^2 \eta} \left[
            \frac{1}{f_2(l-l')} + \frac{1}{f_2(l+l')} \right]
\nn \\
&&+ {\sum_{n,n'=1}^\infty}'
      \frac{(-1)^{l-l'} a_n a_{n'}}
           {2 f_{\frac{n^2}{2\eta}}(2l)f_{\frac{n'^2}{2\eta}}(2l')}
\left\{ \cos\left[q (n l - n' l')\right] 
        \frac{f_{\frac{n n'}{\eta}}(l+l')}
             {f_{\frac{n n'}{\eta}}(l-l')}
      - \cos\left[q (n l + n' l')\right]
        \frac{f_{\frac{n n'}{\eta}}(l-l')}
             {f_{\frac{n n'}{\eta}}(l+l')} 
\right\}
\nn \\
&&- \frac{1}{2\pi \eta} {\sum_{n=1}^\infty}' n a_n 
\left\{ \frac{(-1)^l \cos\left( n q l\right)}
        {f_{\frac{n^2}{2\eta}}(2l)} \left[ g(l+l')+g(l-l')\right]
      + \frac{(-1)^{l'} \cos\left( n q l'\right)}
        {f_{\frac{n^2}{2\eta}}(2l')} \left[ g(l+l')-g(l-l')\right]  \right\},
\label{eq:Czfin}
\eea
\end{widetext}
\be
\langle S^z_l \rangle 
= \frac{q}{2\pi}
   - {\sum_{n=1}^\infty}' a_n 
      \frac{(-1)^l \sin\left( n q l\right)}{f_{n^2/2\eta}(2l)},
\label{eq:Szfin}
\ee
where $q = 2\pi ML/(L+1)$,
\bea
f_\nu(x) &=&
   \left[
\frac{2(L+1)}{\pi}\sin\left(\frac{\pi |x|}{2(L+1)} \right)
\right]^\nu,
   \label{eq:fx} \\
g(x) &=& \frac{\pi}{2(L+1)} \cot\left( \frac{\pi x}{2(L+1)}\right),
          \label{eq:gx}
\eea
and the sum $\sum'$ is taken over odd $n$ only.
We have used the regularization
$\sum^\infty_{n=1}[1-\cos(q_nx)]/n=\ln[f_1(x)]$
as in our previous studies.\cite{corAmp,ladder}
The factor $s(n,n';l,l')$ in Eq.\ (\ref{eq:Cxfin}) is
\bea
&&s(n,n';l,l') 
\nn \\
&&~= \left\{ 
\begin{array}{l}
(-1)^{(n+1)l+(n'+1)l'+(n+n')/2}\\
~~~~~~~~~~~~~~~~~~~~~~~~~~
\mathrm{if}~n + n'= \mathrm{even},\\ \\
(-1)^{(n+1)l+(n'+1)l'+(n'-n+1)/2} \mathrm{sgn}(l-l') \\
~~~~~~~~~~~~~~~~~~~~~~~~~~
\mathrm{if}~n + n'= \mathrm{odd}.
\end{array} 
\right. \nn
\eea
In the thermodynamic limit ($L \to \infty$ with
$|l-L/2| \ll L$, and $|l'-L/2| \ll L$) 
the correlators (\ref{eq:Cxfin}) and (\ref{eq:Czfin}) 
reduce to Eqs.\ (\ref{eq:Cxtdl1}) and (\ref{eq:Cztdl1}).

\subsection{Four-spin correlation functions}

In this subsection, we discuss fusion of two operators
taking $S^x_lS^x_{l+1}$ as an example.
To find bosonic representation of the composite operator
$S^x_lS^x_{l+1}$, we need operator product expansion of the
operators in Eq.\ (\ref{eq:Sx}).
We explicitly write down 
the product of $S^x_l$ and $S^x_{l+1}$ as
\be
S^x_l S^x_{l+1} = \frac{1}{16} 
\sum_{\epsilon_1,\epsilon_2,\epsilon'_1,\epsilon'_2 = \pm 1}
\sum_{n_1,n_2=0}^\infty
X(\{\epsilon_i,\epsilon'_i, n_i\}; l),
\label{eq:SxSx}
\ee
where
\bea
X(\{\epsilon_i,\epsilon'_i, n_i\}; l)
&=& b_{n_1} b_{n_2} t(\{\epsilon_i,\epsilon'_i,n_i;l\})
\nn \\
&&\times \exp\left\{i 2 \pi R [\epsilon_1 \tilde{\phi}(l) 
          + \epsilon_2 \tilde{\phi}(l+1)] \right\}
\nn \\
&&\times \exp\left\{\frac{i}{R} [n_1 \epsilon'_1 \phi(l)
                   + n_2 \epsilon'_2 \phi(l+1) ] \right\},
\nn\\&&
\label{eq:Xcomp}
\eea
and $t(\{\epsilon_i,\epsilon'_i,n_i;l\})$ is defined by 
$t=-1$ ($n_1,n_2 =$ even), 
$t=-\epsilon_1 \epsilon_2 \epsilon'_1 \epsilon'_2$ ($n_1,n_2 =$ odd), 
$t=-i(-1)^l \epsilon_2 \epsilon'_2$ ($n_1=$ even and $n_2 =$ odd), 
$t=i(-1)^l \epsilon_1 \epsilon'_1$ ($n_1=$ odd and $n_2 =$ even).
To find the first few leading operators in the
expansion (\ref{eq:SxSx}),
we make each exponential operator in Eq.\ (\ref{eq:Xcomp})
in normal order and expand the fields $\phi(l+1)$ and 
$\tilde{\phi}(l+1)$ as
$\phi(l+1) = \phi(l) + d\phi(l)/dl + \cdots$. 
It turns out that the leading operators in Eq.\ (\ref{eq:SxSx})
come from the following three contributions:

(i) $\epsilon_1 + \epsilon_2 = 0$ and
 $n_1\epsilon'_1 + n_2\epsilon'_2 = 0$.
In this case
$X$ is expanded as a sum of a constant (identity operator) term,
$d\phi/dx$ and $d\tilde{\phi}/dx$
with scaling dimension 1, and higher-order terms.

(ii) $\epsilon_1 + \epsilon_2 = 0$ and 
$n_1\epsilon'_1 + n_2\epsilon'_2 = \pm 1$.
The leading operators in the expansion of $X$ are
$(-1)^l \exp[ \pm i\phi(l)/R ]$, 
whose scaling dimension is $1/(2\eta)$.

(iii) $\epsilon_1 + \epsilon_2 = \pm 2$ and
$n_1\epsilon'_1 + n_2\epsilon'_2 = 0$.
The leading operators in this case are
$\exp[ \pm i 4\pi R \tilde\phi(l) ]$,
which have dimension $2\eta$.

We thus find that the operator $S^x_lS^x_{l+1}$ has the expansion
\bea
S^x_lS^x_{l+1} 
&=&
 c_0 + c'_0 \frac{d\phi(l)}{dl} + c''_0 \frac{d\tilde{\phi}(l)}{dl}
\nn \\
&&+ c_1 (-1)^l 
\sin \frac{\phi(l)}{R}
+ c'_1 (-1)^l 
\cos \frac{\phi(l)}{R}
\nn \\
&&+ c_2 \cos\left[ 4 \pi R \tilde{\phi}(l) \right] + \cdots.
\label{eq:SxSxboson}
\eea
Apart from the constant term $c_0$, 
the leading term in $S^x_l S^x_{l+1}$ is 
the oscillating one with the coefficients $c_1$ and $c'_1$ 
if $\Delta > 0$ ($\eta > 1/2$), 
while it is the non-oscillating term with $c'_0$ and $c''_0$ 
if $\Delta < 0$ ($\eta < 1/2$).
We note that, for both signs of $\Delta$,
the leading term in the expansion of $S^x_l S^x_{l+1}$
is not simply given by the product of the leading operators
in $S^x_l$ and $S^x_{l+1}$.
In fact, the higher-order terms with large $n$ in Eq.\ (\ref{eq:Sx}),
which can be ignored in the calculation of long-distance asymptotes
of two-point correlation functions,
give contributions to the leading operator
in the operator product expansion of $S^x_lS^x_{l+1}$. 
The same observation can be made for the other subleading terms 
in $S^x_l S^x_{l+1}$.
This result illustrates that,
when one fuses two (or more) operators at short distances,
information on only a few leading terms of 
each operator is, in general, not enough 
to determine the leading terms in the bosonic representation
of the fused operator.
Instead, one must survey contributions from all higher order terms.

Calculating the coefficients $c_n$ from the coefficients $b_n$ is
hardly possible not only because each $c_n$ has contributions from
infinitely many $b_n$s but also because of the ambiguity due to
the short-distance cutoff.
It is thus more practical to estimate the $c_n$s from
the four-spin correlation function
$\langle S^x_lS^x_{l+1} S^x_{l'}S^x_{l'+1}\rangle$ and
its variants.
From Eq.\ (\ref{eq:SxSxboson})
we know the asymptotic behavior of
the four-spin correlation functions, 
\bea
&&
\langle S^x_l S^x_{l+1}S^x_{l'} S^x_{l'+1} \rangle
= B_0 + B_1 \frac{(-1)^{l-l'}}{|l-l'|^{\frac{1}{\eta}}}
 \cos[Q(l-l')] 
\nn \\
&& \hspace*{3cm}
+ \frac{B'_0}{|l-l'|^2} + \frac{B_2}{|l-l'|^{4\eta}} + \cdots,
\label{eq:cx4tdl} \\
&&\langle :\!\!(S^+_l S^-_{l+1} + S^-_l S^+_{l+1})\!\!:\, 
          :\!\!(S^+_{l'} S^-_{l'+1} + S^-_{l'} S^+_{l'+1})\!\!:\rangle 
\nn \\
&&~~= 16 B_1 \frac{(-1)^{l-l'}}{|l-l'|^{\frac{1}{\eta}}}
 \cos[Q(l-l')] 
+ \frac{16 B'_0}{|l-l'|^2} + \cdots,
\qquad
\label{eq:cpmpmtdl} \\
&&\langle S^+_l S^+_{l+1} S^-_{l'} S^-_{l'+1} \rangle
= \frac{8B_2}{|l-l'|^{4\eta}} + \cdots,
\label{eq:cppmmtdl}
\eea
where
$:\!\mathcal{O}\!: = \mathcal{O} - \langle \mathcal{O} \rangle$.
The correlation amplitudes are related to the coefficients $c_n$ by
$B_0 = c_0^2$, $B_1 = (c_1^2 + c_1'^2)/2$, 
$B'_0 = -(c'_0{}^2 + c''_0{}^2)/2\pi$, and $B_2 = c_2^2/2$.

\section{Numerical results}

\begin{figure}
\begin{center}
\noindent
\epsfxsize=0.45\textwidth
\epsfbox{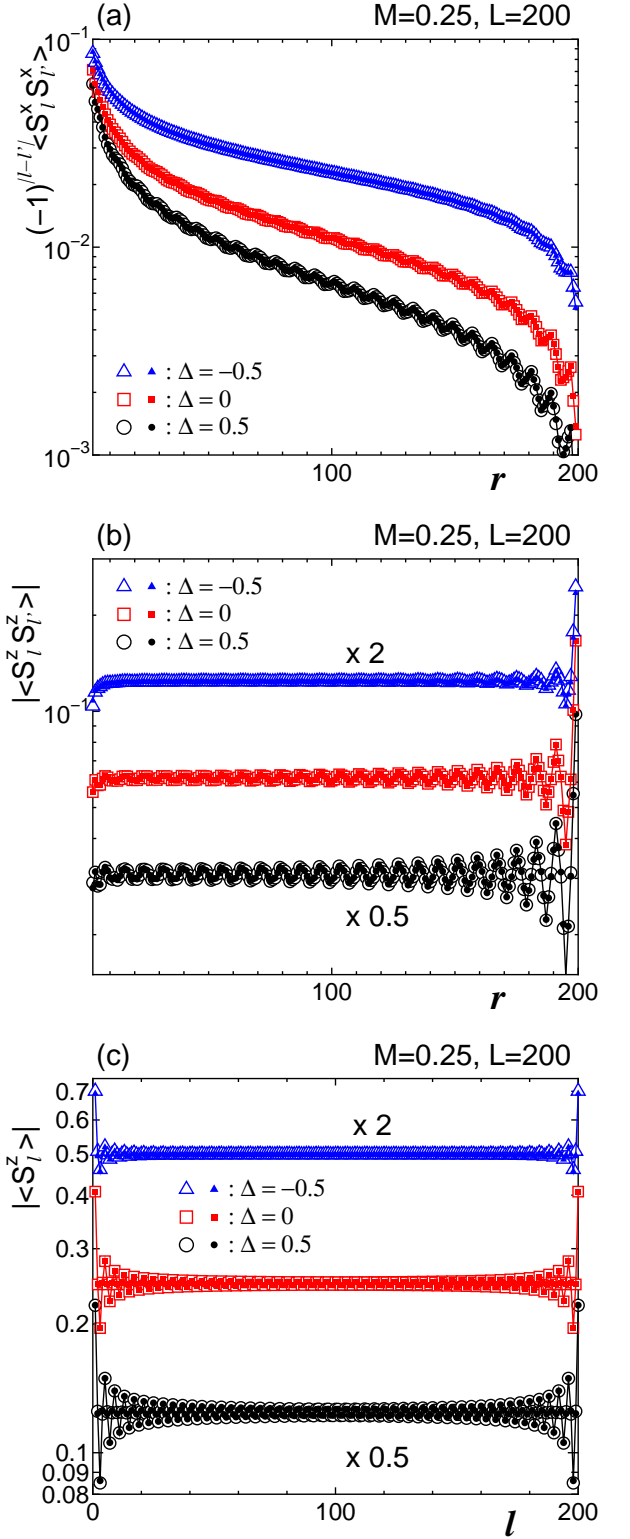}
\end{center}
\caption{
(a) $(-1)^{|l-l'|} \langle S^x_l S^x_{l'} \rangle$ versus $r = |l-l'|$, 
(b) $|\langle S^z_l S^z_{l'}\rangle|$ versus $r$,
(c) $|\langle S^z_l \rangle|$ versus $l$ 
for $\Delta = 0.5$, $0$, and $-0.5$ and $M = 0.25$.
The open symbols are the DMRG data while the small dots are 
the results of fitting.
}
\label{fig:fit}
\end{figure}

In this section, we present numerical results on 
the correlation amplitudes in the two- and four-spin correlation functions 
obtained from the DMRG calculation.\cite{White1,White2}
We calculate the spin polarization $\langle S^z_l \rangle$,
the two-spin correlation functions $\langle S^x_l S^x_{l'} \rangle$
and $\langle S^z_l S^z_{l'} \rangle$, 
and the four-spin correlation functions 
$\langle :\!\!( S^+_l S^-_{l+1} + S^-_{l} S^+_{l+1})\!\!: \,
         :\!\!( S^+_{l'} S^-_{l'+1} + S^-_{l'} S^+_{l'+1})\!\! : \,
 \rangle$
and
$\langle S^+_l S^+_{l+1} S^-_{l'} S^-_{l'+1} \rangle$
in the open $XXZ$ chain of $L = 200$ sites.
The correlation functions are calculated for $l = r_0 - r/2$ and 
$l' = r_0 + r/2$, where $r_0$ represent the center position of the chain, 
$r_0 = L/2$ (for even $r$) or $r_0 = (L+1)/2$ (for odd $r$).
The numerical calculation is done using the finite system algorithm, 
and the number of kept states $m$ is up to 200.
We estimate the numerical error due to the DMRG truncation 
from difference between the data computed with $m = 200$ and
those with $m = 150$.
The estimated errors for the spin polarization, 
two- and four-spin correlation functions are typically 
less than $10^{-7}$, $10^{-6}$, and $10^{-6}$, respectively, 
and sufficiently small for accurate estimation 
of the amplitudes.

\subsection{Amplitudes of two-spin correlation functions}

First, we show the results on the spin polarization
$\langle S^z_l\rangle$ and 
the two-spin correlation functions
$\langle S^x_l S^x_{l'}\rangle$ and $\langle S^z_l S^z_{l'}\rangle$.
Since the $n$th order terms in Eqs.\ (\ref{eq:Sz}) and (\ref{eq:Sx})
contribute to the correlators less and less
for large $n$, we may neglect the higher order terms 
with $n \ge 2$ in the fitting procedure.
That is to say, we fit the DMRG data to the analytic formulas 
(\ref{eq:Cxfin})--(\ref{eq:Szfin}) setting $a_n = b_n = 0$ for $n \ge 2$ 
and taking $b_0$, $b_1$, and $a_1$ as fitting parameters.
We note that this scheme for determining the coefficients 
is basically the same as those used in our previous
studies,\cite{corAmp,ladder} 
in one\cite{ladder} of which the numerical data of $A^z_1$ 
and $A^x_0$ are reported for several typical values of $M$ and
$0 \le \Delta \le 1$.
However, in that work\cite{ladder} the decay exponent $\eta$ as well
as the coefficients was taken as a fitting parameter, and this could
cause small but avoidable errors in the estimates of the coefficients.
In the present work, we use the exact value of $\eta$ obtained from 
the Bethe ansatz solutions.
We therefore believe that the estimates of the coefficients presented
here are even more accurate than the previous ones.

\begin{figure*}
\begin{center}
\noindent
\epsfxsize=0.95\textwidth
\epsfbox{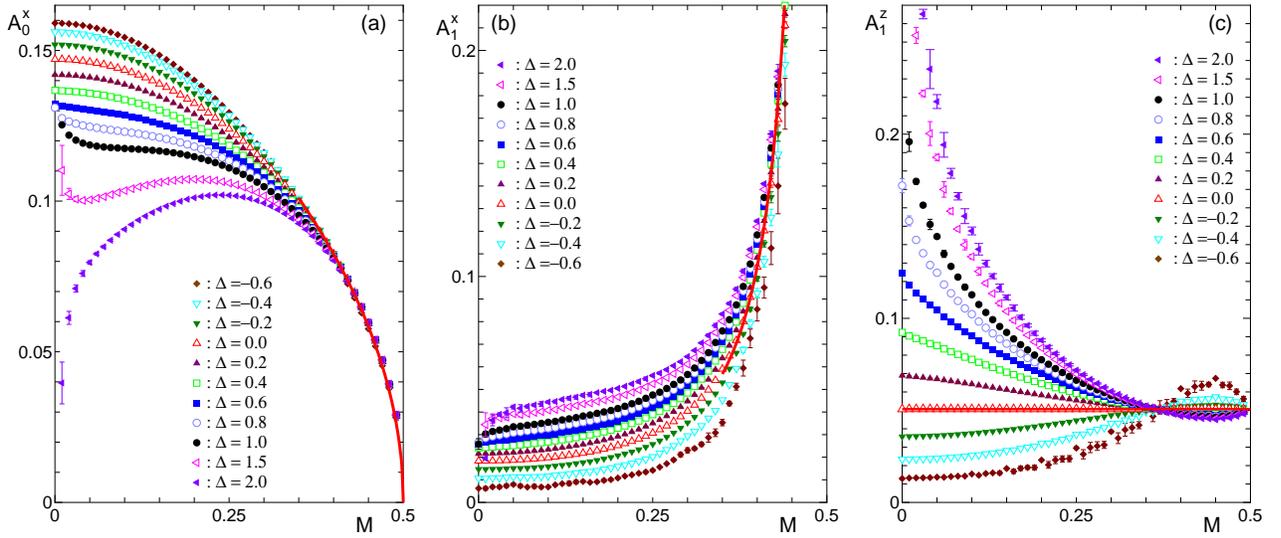}
\end{center}
\caption{
Amplitudes of the two-spin correlation functions as functions of $M$ 
for several typical values of $\Delta$;
(a) $A^x_0$, (b) $A^x_1$, and (c) $A^z_1$.
$A^x_0$ and $A^x_1$ shown in (a) and (b) are estimated from 
$\langle S^x_l S^x_{l'} \rangle$. 
$A^z_1$ for $M > 0$ in (c) are estimated from $\langle S^z_l \rangle$
while $A^z_1$ at $M = 0$ are from $\langle S^z_l S^z_{l'} \rangle$. 
The solid curves in (a) and (b) represent
Eqs.\ (\ref{eq:VT0}) and (\ref{eq:VT1}), respectively, 
while the solid line in (c) represents the exact result
$A^z_1 = 1/(2\pi^2)$ valid at $\Delta = 0$.
}
\label{fig:Amp}
\end{figure*}

Figure \ref{fig:fit} shows the two-spin correlation functions 
$\langle S^x_l S^x_{l'} \rangle$, $\langle S^z_l S^z_{l'} \rangle$, 
and the spin polarization $\langle S^z_l \rangle$ 
at $\Delta = 0.5, 0, -0.5$ and $M = 0.25$.
The DMRG data and the fitting results are plotted 
with the open and solid symbols, respectively.
The excellent agreement between them
demonstrates that the fitting procedure works extremely well
at least for the parameters used in Fig.\ \ref{fig:fit}.
To determine the correlation amplitudes, we perform the fitting 
for the data of several ranges, $20 \le r \le 140$, 
$20 \le r \le 180$, $60 \le r \le 140$, and $60 \le r \le 180$ 
for the two-spin correlation functions, and 
$20 \le l \le 180$, $40 \le l \le 160$, and $60 \le l \le 140$
for the spin polarization.
We then take the mean and the variance of the fitting results 
as the estimate and the error of the numerical values 
of the amplitudes, respectively.
The results of $A^x_0=b_0^2/2$, $A^x_1=b_1^2/4$ 
estimated from $\langle S^x_l S^x_{l'} \rangle$ 
and $A^z_1=a_1^2/2$ estimated from $\langle S^z_l \rangle$ 
are plotted in Fig.\ \ref{fig:Amp} as functions of $M$ 
for several typical values of $\Delta$.
As shown in the figure, the amplitudes take 
non-universal values at $M = 0$ and vary smoothly as $M$ increases.
We have confirmed that the data of $A^z_1$ estimated from the correlation
function $\langle S^z_l S^z_{l'} \rangle$ coincide with
those obtained from $\langle S^z_l \rangle$ within error bars.
We note that, for small $\Delta \alt -0.8$, 
the accuracy of the estimated amplitudes
$A^x_1$ and $A^z_1$ becomes considerably poor. 
This difficulty might be due to the fact that, 
for this parameter range, the subleading terms $\propto A^x_1,A^z_1$
of the correlation functions become considerably smaller than
the non-oscillating leading terms.
Further studies with a more elaborated scheme will be required
for accurate estimation of $A_1^{x,z}$ in this case.

To further examine the accuracy of the numerical estimates obtained above, 
we compare them with exact results which are
available for some limiting cases.

At $M = 0$, the correlation amplitudes $A^x_0$, $A^x_1$ and $A^z_1$ 
for $-1 < \Delta < 1$ 
are analytically calculated:\cite{LukyanovZ,Lukyanov,LukyanovT}
\begin{widetext}
\bea
A^x_0&=&
\frac{1}{8(1-\eta)^2}
\left[\frac{\Gamma(\frac{\eta}{2(1-\eta)})}
            {2\sqrt{\pi}\,\Gamma(\frac{1}{2(1-\eta)})}
\right]^\eta
\exp\left[
-\int^\infty_0\frac{dt}{t}
  \left(\frac{\sinh(\eta t)}{\sinh(t)\cosh[(1-\eta)t]}
        -\eta e^{-2t}\right)\right],
\label{eq:A0xLuky} \\
A^x_1 &=& 
\frac{1}{2\eta(1-\eta)}
\left[\frac{\Gamma(\frac{\eta}{2(1-\eta)})}
            {2\sqrt{\pi}\,\Gamma(\frac{1}{2(1-\eta)})}
\right]^{\eta+\frac{1}{\eta}}
\exp\left[
-\int^\infty_0\frac{dt}{t}
  \left(\frac{\cosh(2\eta t)e^{-2t}-1}{2\sinh(\eta t)\sinh(t)\cosh[(1-\eta)t]}
     +\frac{1}{\sinh(\eta t)}-\frac{\eta^2+1}{\eta} e^{-2t}\right)\right],
\nn \\
\label{eq:A1xLuky} \\
A^z_1 &=&
\frac{2}{\pi^2}
\left[\frac{\Gamma(\frac{\eta}{2(1-\eta)})}
            {2\sqrt{\pi}\,\Gamma(\frac{1}{2(1-\eta)})}
\right]^{\frac{1}{\eta}}
\exp\left[
  \int^\infty_0\frac{dt}{t}
  \left(\frac{\sinh[(2\eta-1) t]}{\sinh(\eta t)\cosh[(1-\eta)t]}
        -\frac{2\eta -1}{\eta} e^{-2t}\right)\right],
\label{eq:A1zLuky}
\eea
\end{widetext}
where $\Gamma(x)$ is the Gamma function.
These equations were previously confirmed by numerical
calculations,\cite{corAmp,ladder,Lukyanov}
and here we have found that the numerical estimates 
of the present work with higher accuracy are even in better agreement
with the above exact formulas for $-0.8 \alt \Delta \alt 0.8$.

In the saturation limit $M \to 1/2$,
exact asymptotic form of the correlation $\langle S^x_l S^x_{l'} \rangle$ 
can be obtained from known exact results on 
the hard-core boson model.\cite{VaidyaT1,VaidyaT2,CreamerTW}
It follows that near the saturation limit the amplitudes $A^x_0$ and
$A^x_1$ should behave as
\bea
A^x_0 &=&
\frac{\rho}{2\sqrt{\pi}} \left( \frac{1}{2}-M \right)^{\frac{1}{2}},
\label{eq:VT0} \\
A^x_1 &=&
\frac{\rho}{16 \pi^{5/2}} 
\left( \frac{1}{2}-M \right)^{-\frac{3}{2}},
\label{eq:VT1}
\eea
where $\rho$ is a universal constant related to Glaisher's constant $A$
by $\rho = \pi e^{1/2} 2^{-1/3} A^{-6} = 0.92418\cdots$.
We can see in Fig.\ \ref{fig:Amp} (a) and (b) that, 
for all $\Delta$s shown, the numerical data of $A^x_0$ and $A^x_1$ 
approach the predicted behavior shown by the solid curves.
Note that there is no free parameter in the theoretical predictions 
(\ref{eq:VT0}) and (\ref{eq:VT1}).

As for the $\langle S^z_l S^z_{l'} \rangle$ correlation, 
it is expected from Eq.\ (\ref{eq:Cztdl1}) that the amplitude 
$A^z_1$ should converge to a universal value $1/(2\pi^2)$ at $M \to 1/2$ 
for arbitrary $\Delta$, since in this limit $\eta$ becomes $1/2$ and
the correlator must take the constant value $1/4$.
Furthermore, for $\Delta = 0$ one can easily calculate
$\langle S^z_lS^z_{l'} \rangle$ exactly using
the Jordan-Wigner transformation
to find $A^z_1 = 1/(2\pi^2)$ for arbitrary $M$.
We clearly see that the numerical data of $A^z_1$ 
in Fig.\ \ref{fig:Amp} (c) agree with these predictions.

From these observations, we conclude that our estimates for the
correlation amplitudes $A^x_0$, $A^x_1$, and $A^z_1$ are pretty accurate,
except for the following parameter regime:
(i) $\Delta\alt -0.8$ and
(ii) $\Delta\agt0.8$ and $M=0$.
In the latter regime the leading irrelevant operator $\cos(2\phi/R)$ 
is no longer negligible.
We point out that diverging behavior at $M=0$, due to the presence of
the (almost) marginal operator, can be clearly seen in the data of 
$A_0^x$ and $A_1^z$ for $\Delta \gtrsim 0.8$.

We summarize the numerical data of
$b_0$, $b_1$, and $a_1$ in Table \ref{tab:Amp}.
It is important to note that the sign of $a_1$ is uniquely determined
once the boundary conditions (\ref{eq:bc}) are fixed.
On the other hand, $b_0$ and $b_1$ have the ambiguity of a common
phase factor.
In the Table, we show the data with fixing
$b_0$ to be real and positive.

\subsection{Amplitudes of four-spin correlation functions}

\begin{figure}
\begin{center}
\noindent
\epsfxsize=0.43\textwidth
\epsfbox{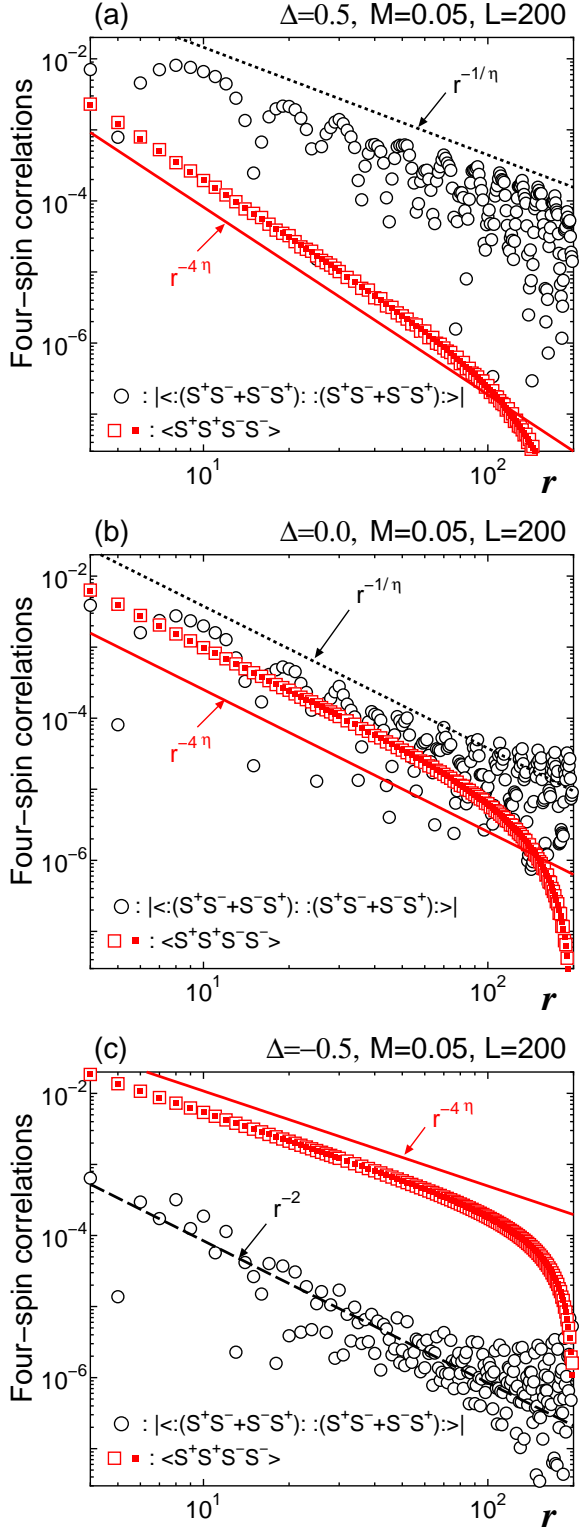}
\end{center}
\caption{
Four-spin correlation functions 
$|\langle :( S^+_l S^-_{l+1} + S^-_{l} S^+_{l+1}): 
          :( S^+_{l'} S^-_{l'+1} + S^-_{l'} S^+_{l'+1}): \rangle |$ 
(open circles) and 
$\langle S^+_l S^+_{l+1} S^-_{l'} S^-_{l'+1} \rangle$ (open squares) 
versus $r = |l-l'|$ for $M = 0.05$ and 
(a) $\Delta = 0.5$, (b) $\Delta = 0$, and (c) $\Delta = -0.5$.
The solid, dotted, and dashed lines correspond, respectively, to 
the algebraic decay of $r^{-4\eta}$, $r^{-1/\eta}$, and $r^{-2}$.
The fitting results for 
$\langle S^+_l S^+_{l+1} S^-_{l'} S^-_{l'+1} \rangle$ 
using Eq.\ (\ref{eq:Cppmmfinite}) are plotted by solid squares.
}
\label{fig:fitcx4}
\end{figure}

\begin{figure}
\begin{center}
\noindent
\epsfxsize=0.4\textwidth
\epsfbox{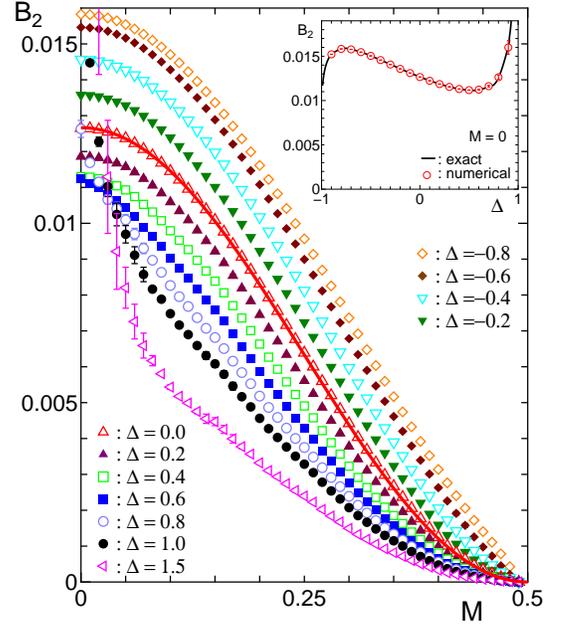}
\end{center}
\caption{
Amplitude $B_2$ of the four-spin correlation function 
$\langle S^+_l S^+_{l+1} S^-_{l'} S^-_{l'+1} \rangle$
as functions of $M$ for several typical values of $\Delta$.
The solid curve represent the relation 
$B_2 = [1+\cos(2\pi M)]/16\pi^2$ expected for $\Delta = 0$.
Inset : Numerical estimates of the amplitude $B_2$ at $M=0$ (open circles)
and the analytical prediction Eq.\ (\ref{eq:B2}) (solid curve).
}
\label{fig:AmpSx4}
\end{figure}

Next we discuss numerical results of
the four-spin correlation functions.
Figure \ref{fig:fitcx4} shows the numerical data of 
the four-spin correlation functions 
$\langle\, : \!(S^+_l S^-_{l+1} + S^-_l S^+_{l+1})\! : \, 
: \!(S^+_{l'} S^-_{l'+1} + S^-_{l'} S^+_{l'+1})\! : \, \rangle$ 
and $\langle S^+_l S^+_{l+1} S^-_{l'} S^-_{l'+1} \rangle$ 
for $\Delta = 0.5$, $0$, and $-0.5$ and $M = 0.05$.
We see that the correlator
$\langle :\!\!(S^+_l S^-_{l+1} + S^-_l S^+_{l+1})\!\! : \,
:\!\!(S^+_{l'} S^-_{l'+1} + S^-_{l'} S^+_{l'+1})\!\! : \rangle$ 
shows power-law decay with the exponent equal to either
$1/\eta$ (for $\Delta \ge 0$) 
or $2$ (for $\Delta < 0$),
while $\langle S^+_l S^+_{l+1} S^-_{l'} S^-_{l'+1} \rangle$ always
decays with the exponent $4\eta$.
We have found the same behavior for other values of $\Delta$ and $M$.
These results are consistent with 
Eqs.\ (\ref{eq:cx4tdl})--(\ref{eq:cppmmtdl}).

To estimate the amplitude $B_2 = c_2^2/2$, 
we fit the numerical data of 
$\langle S^+_l S^+_{l+1} S^-_{l'} S^-_{l'+1} \rangle$ 
to its analytic formula for finite chains with open boundaries,
\bea
&&\langle S^+_l S^+_{l+1} S^-_{l'} S^-_{l'+1} \rangle
\nn \\
&&~~~~~~~~= (2 c_2)^2 \langle \exp\left[ i 4\pi R \tilde{\phi}(l)\right] 
                      \exp\left[-i 4\pi R \tilde{\phi}(l')\right] \rangle
\nn \\
&&~~~~~~~~= 8 B_2 
\frac{f_{2\eta}(2l) f_{2\eta}(2l')}{f_{4\eta}(l+l') f_{4\eta}(l-l')},
\label{eq:Cppmmfinite}
\eea
where we used Eq.\ (\ref{eq:tphi}).
We see in Fig.\ \ref{fig:fitcx4} that the data of 
$\langle S^+_l S^+_{l+1} S^-_{l'} S^-_{l'+1} \rangle$ 
are fitted by the formula extremely well.
The estimated values of $B_2$ from the fitting procedure are shown 
in Fig.\ \ref{fig:AmpSx4}.
We see that for each $\Delta$ the amplitude takes 
a non-universal value at $M=0$, decreases monotonically as $M$ increases, 
and vanishes eventually at $M \to 1/2$.
We note that for $\Delta =0$ the analytic form of 
$\langle S^+_l S^+_{l+1} S^-_{l'} S^-_{l'+1} \rangle$ 
can be easily calculated,  yielding
$B_2 = [1+\cos(2\pi M)]/16\pi^2$.
Figure \ref{fig:AmpSx4} shows that the numerical data
for $\Delta = 0$ are in good agreement with the formula.
We also compare the numerical estimate of $B_2$
at $M=0$ and $-1 < \Delta < 1$ with the exact formula of $B_2$ 
derived recently by Lukyanov and Terras,\cite{LukyanovT}
\be
B_2=\frac{[\Gamma(\eta)]^4}{2^{3+4\eta}\pi^{2+2\eta}(1-\eta)^2}
\left[
\frac{\Gamma(\frac{1}{2-2\eta})}{\Gamma(\frac{\eta}{2-2\eta})}
\right]^{4-4\eta}.
\label{eq:B2}
\ee
Here again we find good agreement between the exact result
and our numerical data.
This is another evidence that our estimates are highly reliable.
At $M=0$ and $\Delta \ge 1$, $B_2$ shows a diverging behavior due to 
the marginal operator, suggesting the break down of our analysis.

Unfortunately, we cannot achieve a precise estimation of 
the amplitude $B_1$ of the leading oscillating term in 
$\langle :\! (S^+_l S^-_{l+1} + S^-_l S^+_{l+1})\! : \,
:\!(S^+_{l'} S^-_{l'+1} + S^-_{l'} S^+_{l'+1})\! : \rangle$ 
due to the presence of subleading terms which give
sizable contributions to the correlation function.
This issue of estimating $B_1$ is left for future studies.

\subsection{Spin gap}

The data of the correlation amplitudes obtained in the preceding
subsections is useful for analyzing effects of perturbations
of single-spin type and exchange-coupling type
in the bosonization framework.
To illustrate how this scheme works, we compute
spin gaps induced by such perturbations to the Hamiltonian (\ref{eq:Ham}).

As an example of the perturbation of the single-spin type, 
we consider effects of the staggered transverse field.
The perturbation to the Hamiltonian (\ref{eq:Ham}) is given by
\be
\mathcal{H}' = - h_s \sum_l (-1)^l S^x_l.
\ee
It has been shown that $\mathcal{H}'$ induces a spin gap. 
This field-induced gap is believed to be the origin of the spin-gap behavior 
observed in Cu benzoate\cite{OshikawaA,AffleckO,EsslerT,Essler,Dender1}
and Yb$_4$As$_3$\cite{Reinders,OshikawaU,Khogi} 
under a uniform field, in which the staggered field emerges 
due to the alternating $g$-tensor and the Dzyaloshinskii-Moriya
interaction.
In these materials, exchange anisotropy is negligibly small and 
the staggered transverse field $h_s$ is proportional to 
the uniform field $H$.
Thus, we may set $\Delta = 1$ and $h_s = \gamma H$, 
where $\gamma$ is a constant specific to each material.

\begin{figure}
\begin{center}
\noindent
\epsfxsize=0.40\textwidth
\epsfbox{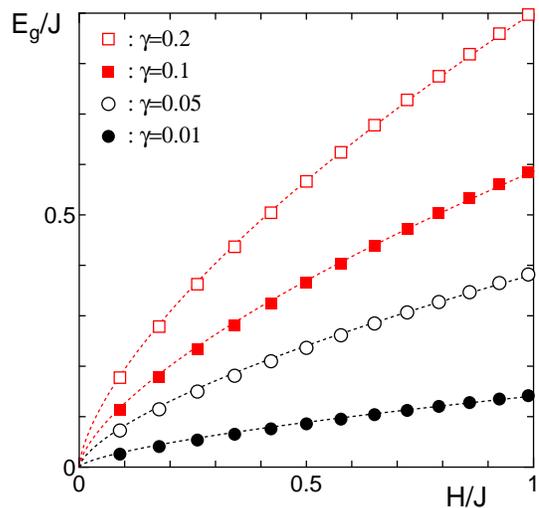}
\end{center}
\caption{
Field-dependence of the spin gap in the Heisenberg chain 
$\mathcal{H}_0(\Delta = 1)$ in a uniform field $H$ and 
a staggered transverse field $h_s = \gamma H$.
The dotted curves represent the expected power-law behavior of the gap, 
$E_g \propto H^{2/3}$.
}
\label{fig:gapstag}
\end{figure}

In the bosonization scheme, the leading uniform term of 
the perturbing Hamiltonian $\mathcal{H}'$, which is responsible 
for opening the gap, is a cosine term,
\be
\widetilde{\mathcal{H}}' = -h_s b_0(H) 
\int \cos\left[ 2 \pi R \tilde{\phi}(x) \right] dx.
\ee
Effects of the perturbation have been studied in the 
literature\cite{OshikawaA,AffleckO,EsslerT,Essler,EsslerFH,Zamolodochikov} 
and the induced spin gap is given by\cite{Zamolodochikov} 
\be
\frac{E_g}{J} = \frac{2 v(H)}{\sqrt{\pi}}
\frac{\Gamma(\frac{\eta}{8-2\eta})}
{\Gamma(\frac{2}{4-\eta})}
\left[\frac{\pi b_0(H)}{2 v(H)}
\frac{\Gamma(\frac{4-\eta}{4})}{\Gamma(\frac{\eta}{4})}
\frac{h_s}{J}\right]^{\frac{2}{4-\eta}}.
\label{eq:gapstag}
\ee
Hence, the field-dependence of the spin gap is evaluated by 
substituting our numerical estimates for $b_0(H)$ as well as 
the exact values of $v(H)$ and $\eta(H)$ into Eq.\ (\ref{eq:gapstag}).
The result is shown in Fig.\ \ref{fig:gapstag} for 
several typical values of $\gamma$.
It reproduces the peculiar $H$-dependence of 
the spin gap observed in experiments, $E_g \sim H^{2/3}$ for small $H$.

\begin{figure}
\begin{center}
\noindent
\epsfxsize=0.4\textwidth
\epsfbox{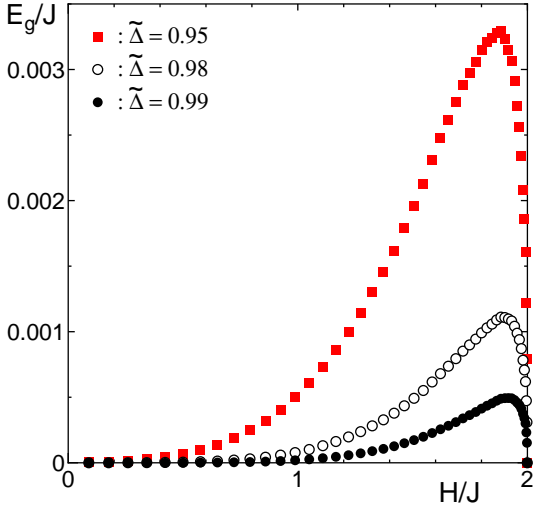}
\end{center}
\caption{
Field-dependence of the spin gap in the Heisenberg chain 
$\mathcal{H}_0(\Delta = 1)$ with the perturbation of 
transverse exchange anisotropy ($1-\widetilde{\Delta}$).
}
\label{fig:gaptrans}
\end{figure}

Next, we consider the spin gap induced by the perturbation 
of exchange anisotropy, 
\be
\mathcal{H}'' = -J(1 - \widetilde{\Delta}) \sum_l S^x_l S^x_{l+1},
\ee
to the Heisenberg chain $\mathcal{H}_0(\Delta=1)$,
where $1-\widetilde\Delta\ll1$.
Note that, by rotating the system around the $y$-axis, 
the whole Hamiltonian is rewritten as
\bea
\mathcal{H}_0 + \mathcal{H}'' &=& 
J \sum_l \left( S^x_l S^x_{l+1} + S^y_l S^y_{l+1} 
         + \widetilde{\Delta} S^z_l S^z_{l+1} \right)
\nn \\
&&
-H \sum_l S^x_l.
\eea
Hence, the system can be also viewed as the $S=1/2$ $XXZ$ chain
with anisotropy 
$\widetilde{\Delta}$ in a uniform transverse field $H$.
As we have seen in Eq.\ (\ref{eq:SxSxboson}),
apart from a constant, the leading operators in the bosonized
$\mathcal{H}''$ are $\frac{d\phi}{dx}$ and $\frac{d\tilde{\phi}}{dx}$.
Since their main effect is just a small renormalization of 
the TL liquid parameter $R$, we can neglect these operators
in lowest order in $1-\widetilde\Delta$.
We thus find that the dominant component which is responsible 
for opening the gap is the cosine term with coefficient $c_2$, 
\be
\widetilde{\mathcal{H}}^{\prime \prime} =
- ( 1 - \widetilde{\Delta}) c_2(H) 
\int \cos\left[ 4 \pi R \tilde{\phi}(x) \right] dx .
\ee
The effect of this perturbation has been studied,\cite{DmitrievKO,CauxEL} 
and the spin gap for 
$H \gg (1-\widetilde{\Delta})$ is found to be\cite{CauxEL}
\be
\frac{E_g}{J} = \frac{2 v(H)}{\sqrt{\pi}}
\frac{\Gamma(\frac{\eta}{2-2\eta})}
{\Gamma(\frac{1}{2-2\eta})}
\left[\frac{\pi(1-\widetilde{\Delta})c_2(H)}{2 v(H)}
\frac{\Gamma(1-\eta)}{\Gamma(\eta)}\right]^{\frac{1}{2-2\eta}}. 
\label{eq:gaptrans}
\ee
We show in Fig.\ \ref{fig:gaptrans} the field dependence of the spin gap 
for several typical values of $\widetilde{\Delta}$.
As shown in the figure, the spin gap opens very slowly with $H$
and closes at the saturation field $H = 2 J$, reflecting the fact 
that the coefficient $c_2$ vanishes as $M \to 1/2$.
In contrast to the case of the staggered transverse field, 
the gap induced by the exchange anisotropy is extremely small.
This result is consistent with the observation of the recent numerical 
study\cite{CapraroG} which finds no substantial gap;
the spin gap is too small to be detected by the numerical study on 
finite-size systems.
We note that the bosonization scheme with our estimates of $c_2$ is,
at present, the only way to get reliable quantitative results 
on the spin gap behavior.

\section{Summary}
In this paper, we have studied the ground-state correlation functions 
in the $S=1/2$ $XXZ$ chain in a magnetic field.
With the bosonic representation of the spin operators $S^z_l$ and
$S^x_l$ for open boundary conditions,
we have calculated the two- and four-spin correlation functions
analytically within the effective low-energy theory.
We have also calculated the correlation functions numerically 
using the DMRG method, and estimated correlation amplitudes 
of the first few leading terms by fitting the numerical results 
to the analytic formulas.
We have thus obtained precise data of non-universal coefficients 
appearing in the bosonic representation of lattice spin operators.
Excellent agreement is found in the
comparison of the numerical data with
the exact known results in various limiting cases.

We believe that the data of the correlation amplitudes presented
in this work is suitable for quantitatively studying low-energy
properties of perturbed spin chains within the bosonization method.
Indeed, it has been shown in Ref.\ \onlinecite{EsslerFH} that 
the data of $a_1$ and $b_0$ can be used successfully to 
explain quantitative features of the dynamical spin structure factor
in Cu benzoate.
We hope that the data will be applied to a wider variety of problems
in one-dimensional spin systems.

\acknowledgments
The authors appreciate F.\ H.\ L.\ Essler 
for fruitful discussions and comments.
The work of A.F.\ was supported in part by a Grant-in-Aid for
Scientific Research on Priority Areas from the Ministry of Education,
Culture, Sports, Science and Technology, Japan
(Grant No.\ 12046238).

\appendix
\section{Bosonic representation of spin operators}

In this appendix, we briefly overview the derivation of 
the bosonic representation of the spin operators 
(\ref{eq:Sz}) and (\ref{eq:S+}) in the spin-1/2 $XXZ$ chain.
We basically follow the strategy of 
Refs.\ \onlinecite{Affleck} and \onlinecite{SheltonNT}; 
We first bosonize the repulsive Hubbard chain at half filling,
in which charge excitations have a Mott-Hubbard gap.
We then obtain the spin operators
by throwing the gapped charge mode away, 
and generalize the results to the anisotropic $XXZ$ case.

Let us begin with the bosonic representation of electron operators
$(\sigma=\uparrow,\downarrow)$,
\bea
\psi_\sigma(x) &=& \psi_{R,\sigma}(x) + \psi_{L,\sigma}(x),
\label{eq:psie} \\
\psi_{R,\sigma}(x) &=& \frac{\kappa_\sigma}{\sqrt{2\pi a}}
e^{i \sqrt{4\pi}\varphi_{R,\sigma}(x) + ik_{F\sigma}x},
\label{eq:psieR} \\
\psi_{L,\sigma}(x) &=& \frac{\kappa_\sigma}{\sqrt{2\pi a}}
e^{-i \sqrt{4\pi}\varphi_{L,\sigma}(x) - ik_{F\sigma}x},
\label{eq:psieL}
\eea
where $\kappa_\sigma$ are Klein factors obeying 
$\{ \kappa_\sigma, \kappa_{\sigma'} \} = 2\delta_{\sigma,\sigma'}$
and the bosonic fields $\varphi$ obey the commutation
relations,
\bea
&&[\varphi_{R,\sigma}(x), \varphi_{R,\sigma'}(y)] 
= (i/4)\delta_{\sigma,\sigma'} {\rm sgn}(x-y),
\\ 
&&
[\varphi_{L,\sigma}(x), \varphi_{L,\sigma'}(y)] 
= - (i/4)\delta_{\sigma,\sigma'} {\rm sgn}(x-y),
\\
&&
[\varphi_{R,\sigma}(x), \varphi_{L,\sigma'}(y)] 
= - (i/4)\delta_{\sigma,\sigma'}.
\eea
The Fermi wavenumbers $k_{F\sigma}$ are functions of
the magnetization $M$,
$k_{F\upa} = \pi(1/2+M)$ and $k_{F\dwa} = \pi(1/2 - M)$.
We introduce fields $\phi$ and $\tilde{\phi}$ given by 
\bea
\phi_\sigma(x) &=& \varphi_{L,\sigma}(x) + \varphi_{R,\sigma}(x),
\\
\tilde{\phi}_\sigma(x) &=& \varphi_{L,\sigma}(x) - \varphi_{R,\sigma}(x),
\eea
which satisfy 
$[\phi_\sigma(x), \tilde{\phi}_\sigma(y)] 
= - (i/2) [1 + {\rm sgn}(x-y)] \delta_{\sigma, \sigma'}$.
It then follows that 
the electron density becomes
\bea
\rho_\sigma &=& 
\psi^\dagger_{R,\sigma} \psi_{R,\sigma} 
+ \psi^\dagger_{L,\sigma} \psi_{L,\sigma}
\nn \\
&=&
\frac{k_{F\sigma}}{\pi} + \frac{1}{\sqrt{\pi}} \frac{d\phi_\sigma}{dx}.
\eea
The uniform charge and spin densities are
\bea
\rho_c &=& \frac{1}{2} (\rho_\upa + \rho_\dwa) 
= \frac{1}{2} + \frac{1}{\sqrt{2\pi}} \frac{d\phi_c}{dx},
\\
\rho_s &=& \frac{1}{2} (\rho_\upa - \rho_\dwa) 
= M + \frac{1}{\sqrt{2\pi}} \frac{d\phi_s}{dx},
\eea
where the charge and spin fields are defined by
\bea
\phi_{c} &=& \frac{1}{\sqrt{2}}(\phi_\upa + \phi_\dwa),
\qquad
\phi_{s} = \frac{1}{\sqrt{2}}(\phi_\upa - \phi_\dwa),
\\
\tilde{\phi}_{c} &=& 
\frac{1}{\sqrt{2}}(\tilde{\phi}_\upa + \tilde{\phi}_\dwa),
\qquad
\tilde{\phi}_{s} = 
\frac{1}{\sqrt{2}}(\tilde{\phi}_\upa - \tilde{\phi}_\dwa).
\eea

The charge mode in the Hubbard chain is gapped at half filling
when the on-site interaction is repulsive, $U > 0$.
The charge gap is generated by the Umklapp scattering term, 
\bea
&&U \left[ \psi^\dagger_{R, \upa}(x)\psi_{L, \upa}(x)
         \psi^\dagger_{R, \dwa}(x)\psi_{L, \dwa}(x) \right.
\nn \\
&&\left.~~~+ \psi^\dagger_{L, \upa}(x)\psi_{R, \upa}(x)
         \psi^\dagger_{L, \dwa}(x)\psi_{R, \dwa}(x) \right]
\nn \\
&&~~~= - \frac{2U}{(2\pi a)^2} \cos(\sqrt{8\pi}\phi_c),
\eea
which pins the charge field
at $\phi_c=n\sqrt{\pi/2}$ ($n$: integer).
At low energies we may treat the field as a classical number, i.e., 
\bea
\cos(\sqrt{8\pi}\phi_c) = C,
\nn \\
\sin(\sqrt{8\pi}\phi_c) = 0,
\eea
where $C$ is a positive non-universal constant.
At this point, following Ref. \onlinecite{Haldane}, we modify 
Eqs.\ (\ref{eq:psieR}) and (\ref{eq:psieL})
to 
\bea
\psi_{R,\sigma}(x) &=& \frac{\kappa_\sigma}{\sqrt{2\pi a}}
\sum_{n=0}^\infty 
e^{i (2n+1) (k_{F\sigma}x + \sqrt{\pi}\phi_\sigma)
             -i\sqrt{\pi}\tilde{\phi}_\sigma},
\nn \\
\psi_{L,\sigma}(x) &=& \frac{\kappa_\sigma}{\sqrt{2\pi a}}
\sum_{n=0}^\infty 
e^{-i (2n+1) (k_{F\sigma}x+\sqrt{\pi}\phi_\sigma)
             -i\sqrt{\pi}\tilde{\phi}_\sigma}.
\nn
\eea

Using the equations above, one can derive the bosonic representation
for the spin operators.
The $z$-component of the spin operator is given by 
\bea
&&S^z(x) 
\nn \\
&&~~= \frac{1}{2} \left[\psi^\dagger_\upa(x) \psi_\upa(x) 
                    - \psi^\dagger_\dwa(x) \psi_\dwa(x) \right]
\nn \\
&&~~= \frac{1}{2} \left[ \rho_s(x) 
+ \psi^\dagger_{R,\upa}(x) \psi_{L,\upa}(x)
+ \psi^\dagger_{L,\upa}(x) \psi_{R,\upa}(x) \right.
\nn \\
&&\left.~~~~~~~~~
- \psi^\dagger_{R,\dwa}(x) \psi_{L,\dwa}(x)
- \psi^\dagger_{L,\dwa}(x) \psi_{R,\dwa}(x)
\right]
\nn \\
&&~~= M + \frac{1}{\sqrt{2\pi}} \frac{d\phi_s}{dx} 
\nn \\
&&~~~- \sum_{n=0}^\infty a_{2n+1}(-1)^x 
\sin\left[ (2n+1)(2\pi Mx + \sqrt{2\pi}\phi_s)\right],
\nn \\
\label{eq:SzHub}
\eea
where $a_{2n+1}$ is a non-universal constant.
Here we must recall that Eq.\ (\ref{eq:SzHub}) is obtained from 
the Hubbard chain at half filling, whose low-energy effective
spin Hamiltonian is nothing but
the antiferromagnetic Heisenberg spin chain,
in which $R = 1/\sqrt{2\pi}$.
To generalize the result to the $XXZ$ chain, what one needs to do
is replacing $\sqrt{2\pi}\phi_s$ with $\phi_s/R$.
We also define $\phi(x)=\phi_s(x)+2\pi RMx$ to obtain
\bea
S^z(x) 
&=& \frac{1}{2\pi R} \frac{d\phi(x)}{dx} 
\nn \\
&&- {\sum_{n=0}^\infty} a_{2n+1} (-1)^x 
\sin\left[ (2n+1)\frac{\phi(x)}{R} \right].
\eea
Similarly, the operator $S^+$ in the antiferromagnetic Heisenberg chain
is given by 
\bea
&&S^+(x) 
\nn \\
&&~~= \left[ \psi^\dagger_{R,\upa}(x) + \psi^\dagger_{L,\upa}(x)\right]
           \left[ \psi_{R,\dwa}(x) + \psi_{L,\dwa}(x) \right]
\nn \\
&&~~= e^{i \sqrt{2\pi}\tilde{\phi}_s}
\sum_{n=0}^\infty \left\{ 
b_{2n} (-1)^x \cos\left[ 2n (2\pi Mx + \sqrt{2\pi}\phi_s)\right] \right.
\nn \\
&&\left.
~~~~~~~~+ b_{2n+1} \sin\left[ (2n+1) (2\pi Mx + \sqrt{2\pi}\phi_s)\right]
\right\}.
\nn \\
\eea
By generalizing the equation to the $XXZ$ case, 
we arrive at the final formula, 
\bea
&&S^+(x) 
\nn \\
&&~~= e^{i 2\pi R \tilde{\phi}}
\sum_{n=0}^\infty \left\{ 
b_{2n} (-1)^x \cos\left[ 2n \frac{\phi(x)}{R}\right]
\right.
\nn \\
&&\left.~~~~~~~~+ b_{2n+1} 
\sin\left[ (2n+1) \frac{\phi_s(x)}{R}\right] \right\},
\nn \\
\eea
where we have replaced $\sqrt{2\pi} \tilde{\phi}_s$ with
$2\pi R \tilde{\phi}$.

\newpage

\begin{table*}
\caption{
\label{tab:Amp}
Amplitudes (a) $b_0$, and (b) $b_1$, and (c) $a_1$
as functions of the magnetization $M$ for several typical values of $\Delta$.
The values of $a_1$ in (c) for $M > 0$ are estimated 
from the fitting of $\langle S^z_l \rangle$ 
while those for $M = 0$ are from $\langle S^z_l S^z_{l'} \rangle$.
The figures in parenthesis indicate the estimated error 
on the last quoted digits.
The error of the data without parenthesis 
is estimated to be less than $10^{-4}$.
The exact values of 
Eqs. (\ref{eq:A0xLuky})--(\ref{eq:A1zLuky}) for $M = 0$ are also listed.
}
(a) $b_0$ ($1 \ge \Delta > 0$)
\begin{ruledtabular}
\begin{tabular}{ccccccccccc}
 $M$ & $\Delta=1.0$ & $\Delta=0.9$ & $\Delta=0.8$ & $\Delta=0.7$ 
 & $\Delta=0.6$ & $\Delta=0.5$ & $\Delta=0.4$ & $\Delta=0.3$ & 
 $\Delta=0.2$ & $\Delta=0.1$ \\
\hline
0.00(exact) 
&   *****
&  0.5141
&  0.5079
&  0.5094
&  0.5131
&  0.5177
&  0.5226
&  0.5277
&  0.5327
&  0.5376
\\
\hline
   0.00
 &       ***** 
 &      0.521(4) 
 &      0.512(2) 
 &      0.512(1) 
 &     0.5144(6) 
 &     0.5184(3) 
 &     0.5230(1) 
 &     0.5279
 &     0.5328
 &     0.5376
  \\
   0.01
 &     0.5006(3) 
 &     0.5022
 &     0.5050(1) 
 &     0.5088(1) 
 &     0.5131(1) 
 &     0.5178(1) 
 &     0.5227
 &     0.5277
 &     0.5327
 &     0.5376
  \\
   0.02
 &     0.4937(2) 
 &     0.4983
 &     0.5029
 &     0.5076
 &     0.5125
 &     0.5174
 &     0.5225
 &     0.5275
 &     0.5325
 &     0.5374
  \\
   0.03
 &     0.4904(2) 
 &     0.4960
 &     0.5014
 &     0.5067
 &     0.5119
 &     0.5170
 &     0.5221
 &     0.5272
 &     0.5322
 &     0.5371
  \\
   0.04
 &     0.4883(2) 
 &     0.4945(1) 
 &     0.5003
 &     0.5058
 &     0.5112
 &     0.5165
 &     0.5217
 &     0.5268
 &     0.5318
 &     0.5366
  \\
   0.05
 &     0.4871(2) 
 &     0.4934(1) 
 &     0.4994
 &     0.5051
 &     0.5106
 &     0.5159
 &     0.5212
 &     0.5262
 &     0.5312
 &     0.5360
  \\
\end{tabular}
\end{ruledtabular}
\end{table*}

\newpage
\begin{table*}
(a) $b_0$ ($0 \ge \Delta \ge -0.6, \Delta = 1.5, 2.0$)
\begin{ruledtabular}
\begin{tabular}{cccccccc|cc}
 $M$ & $\Delta=0.0$ & $\Delta=-0.1$ & $\Delta=-0.2$ & $\Delta=-0.3$ 
 & $\Delta=-0.4$ & $\Delta=-0.5$ & $\Delta=-0.6$ & $\Delta=1.5$ & 
 $\Delta=2.0$ \\
\hline
0.00(exact)
&  0.5424
&  0.5470
&  0.5513
&  0.5553
&  0.5589
&  0.5619
&  0.5641
&  *****
&  *****
\\
\hline
   0.00
 &     0.5424
 &     0.5469
 &     0.5512
 &     0.5552
 &     0.5588
 &     0.5618
 &     0.5640
 &     *****
 &     *****
  \\
   0.01
 &     0.5423
 &     0.5469
 &     0.5512
 &     0.5551
 &     0.5587
 &     0.5617
 &     0.5640
 &       0.47(1) 
 &       0.28(2) 
  \\
   0.02
 &     0.5421
 &     0.5467
 &     0.5509
 &     0.5549
 &     0.5585
 &     0.5615
 &     0.5637
 &      0.453(4) 
 &      0.350(6) 
  \\
   0.03
 &     0.5418
 &     0.5463
 &     0.5506
 &     0.5545
 &     0.5581
 &     0.5611
 &     0.5633
 &      0.449(1) 
 &      0.377(3) 
  \\
   0.04
 &     0.5413
 &     0.5458
 &     0.5501
 &     0.5540
 &     0.5575
 &     0.5605
 &     0.5627
 &      0.448(1) 
 &      0.390(2) 
  \\
   0.05
 &     0.5407
 &     0.5452
 &     0.5494
 &     0.5533
 &     0.5568
 &     0.5598
 &     0.5620
 &     0.4482(9) 
 &      0.399(1) 
  \\
\end{tabular}
\end{ruledtabular}
\end{table*}

\newpage
\begin{table*}
(b) $b_1$ ($1 \ge \Delta > 0$)
\begin{ruledtabular}
\begin{tabular}{ccccccccccc}
 $M$ & $\Delta=1.0$ & $\Delta=0.9$ & $\Delta=0.8$ & $\Delta=0.7$ 
 & $\Delta=0.6$ & $\Delta=0.5$ & $\Delta=0.4$ & $\Delta=0.3$ & 
 $\Delta=0.2$ & $\Delta=0.1$ \\
\hline
0.00(exact)
&  *****
&  0.3321
&  0.3323
&  0.3297
&  0.3254
&  0.3195
&  0.3124
&  0.3040
&  0.2943
&  0.2834
\\
\hline
   0.00
 &       *****
 &      0.325(5) 
 &      0.326(2) 
 &      0.325(1) 
 &     0.3222(4) 
 &     0.3173(1) 
 &     0.3109
 &     0.3031
 &     0.2938
 &     0.2832
  \\
   0.01
 &      0.348(7) 
 &      0.343(5) 
 &      0.338(3) 
 &      0.332(2) 
 &      0.326(1) 
 &      0.320(1) 
 &     0.3121(7) 
 &     0.3036(5) 
 &     0.2941(3) 
 &     0.2833(1) 
  \\
   0.02
 &      0.355(3) 
 &      0.348(2) 
 &      0.342(1) 
 &      0.336(1) 
 &     0.3287(8) 
 &     0.3214(6) 
 &     0.3133(4) 
 &     0.3044(3) 
 &     0.2946(2) 
 &     0.2837(1) 
  \\
   0.03
 &     0.3580(9) 
 &     0.3516(9) 
 &     0.3450(7) 
 &     0.3382(6) 
 &     0.3309(5) 
 &     0.3232(3) 
 &     0.3147(2) 
 &     0.3055(1) 
 &     0.2954
 &     0.2843
  \\
   0.04
 &      0.361(2) 
 &      0.355(1) 
 &      0.348(1) 
 &      0.341(1) 
 &     0.3329(9) 
 &     0.3247(8) 
 &     0.3159(6) 
 &     0.3065(5) 
 &     0.2962(4) 
 &     0.2850(2) 
  \\
   0.05
 &      0.366(1) 
 &     0.3588(9) 
 &     0.3513(8) 
 &     0.3434(7) 
 &     0.3352(6) 
 &     0.3266(5) 
 &     0.3174(4) 
 &     0.3077(4) 
 &     0.2972(3) 
 &     0.2859(2) 
  \\
\end{tabular}
\end{ruledtabular}
\end{table*}

\newpage

\begin{table*}
(b) $b_1$ ($0 \ge \Delta \ge -0.6, \Delta = 1.5, 2.0$)
\begin{ruledtabular}
\begin{tabular}{cccccccc|cc}
 $M$ & $\Delta=0.0$ & $\Delta=-0.1$ & $\Delta=-0.2$ & $\Delta=-0.3$ 
 & $\Delta=-0.4$ & $\Delta=-0.5$ & $\Delta=-0.6$ & $\Delta=1.5$ & 
 $\Delta=2.0$ \\
\hline
0.00(exact)
&  0.2712
&  0.2576
&  0.2424
&  0.2255
&  0.2068
&  0.1861
&  0.1630
&  *****
&  *****
\\
\hline
   0.00
 &     0.2712
 &     0.2576
 &     0.2425
 &     0.2255(1) 
 &     0.2063(4) 
 &     0.1844(9) 
 &      0.158(2) 
 &     *****
 &     *****
  \\
   0.01
 &     0.2712
 &     0.2577(1) 
 &     0.2425(3) 
 &     0.2255(7) 
 &      0.206(1) 
 &      0.184(4) 
 &       0.16(1) 
 &       0.37(2) 
 &      0.279(4) 
  \\
   0.02
 &     0.2716
 &     0.2581(1) 
 &     0.2432(3) 
 &     0.2266(7) 
 &      0.208(1) 
 &      0.188(2) 
 &      0.166(7) 
 &       0.39(1) 
 &       0.38(1) 
  \\
   0.03
 &     0.2721
 &     0.2585(1) 
 &     0.2435(3) 
 &     0.2269(7) 
 &      0.209(1) 
 &      0.188(3) 
 &      0.166(7) 
 &     0.3867(1) 
 &      0.395(3) 
  \\
   0.04
 &     0.2728
 &     0.2593(2) 
 &     0.2446(6) 
 &      0.228(1) 
 &      0.211(2) 
 &      0.191(4) 
 &      0.171(8) 
 &      0.388(6) 
 &       0.40(1) 
  \\
   0.05
 &     0.2737
 &     0.2603(1) 
 &     0.2458(5) 
 &      0.230(1) 
 &      0.213(1) 
 &      0.195(3) 
 &      0.179(6) 
 &      0.396(2) 
 &      0.411(4) 
  \\
\end{tabular}
\end{ruledtabular}
\end{table*}

\newpage
\begin{table*}
(c) $a_1$ ($1 \ge \Delta > 0$)
\begin{ruledtabular}
\begin{tabular}{ccccccccccc}
 $M$ & $\Delta=1.0$ & $\Delta=0.9$ & $\Delta=0.8$ & $\Delta=0.7$ 
 & $\Delta=0.6$ & $\Delta=0.5$ & $\Delta=0.4$ & $\Delta=0.3$ & 
 $\Delta=0.2$ & $\Delta=0.1$ \\
\hline
0.00(exact)
&  *****
&  0.7049
&  0.6069
&  0.5464
&  0.5008
&  0.4629
&  0.4297
&  0.3995
&  0.3712
&  0.3444
\\
\hline
   0.00
 &       *****
 &       0.64(1) 
 &      0.587(6) 
 &      0.540(2) 
 &     0.4990(5) 
 &     0.4626(1) 
 &     0.4297(4) 
 &     0.3995(3) 
 &     0.3713(2) 
 &     0.3443(1) 
  \\
   0.01
 &      0.626(8) 
 &      0.589(6) 
 &      0.553(4) 
 &      0.519(3) 
 &      0.486(2) 
 &      0.455(1) 
 &      0.425(1) 
 &     0.3971(8) 
 &     0.3701(4) 
 &     0.3439(2) 
  \\
   0.02
 &      0.590(2) 
 &      0.562(2) 
 &      0.534(1) 
 &      0.505(1) 
 &     0.4771(7) 
 &     0.4493(5) 
 &     0.4220(3) 
 &     0.3953(2) 
 &     0.3692(1) 
 &     0.3436
  \\
   0.03
 &      0.568(1) 
 &      0.545(1) 
 &      0.520(1) 
 &     0.4953(9) 
 &     0.4699(7) 
 &     0.4444(5) 
 &     0.4189(3) 
 &     0.3935(1) 
 &     0.3683(1) 
 &     0.3432
  \\
   0.04
 &      0.549(5) 
 &      0.529(4) 
 &      0.508(3) 
 &      0.486(2) 
 &      0.463(2) 
 &      0.439(1) 
 &      0.415(1) 
 &     0.3913(8) 
 &     0.3670(5) 
 &     0.3427(2) 
  \\
   0.05
 &     0.5375(6) 
 &     0.5190(4) 
 &     0.4995(2) 
 &     0.4790(1) 
 &     0.4577
 &     0.4356
 &     0.4129
 &     0.3897
 &     0.3662
 &     0.3424
  \\
\end{tabular}
\end{ruledtabular}
\end{table*}

\newpage
\begin{table*}
(c) $a_1$ ($0 \ge \Delta \ge -0.6, \Delta = 1.5, 2.0$)
\begin{ruledtabular}
\begin{tabular}{cccccccc|cc}
 $M$ & $\Delta=0.0$ & $\Delta=-0.1$ & $\Delta=-0.2$ & $\Delta=-0.3$ 
 & $\Delta=-0.4$ & $\Delta=-0.5$ & $\Delta=-0.6$ & $\Delta=1.5$ & 
 $\Delta=2.0$  \\
\hline
0.00(exact)
&  0.3183
&  0.2928
&  0.2674
&  0.2418
&  0.2160
&  0.1895
&  0.1621
&  *****
&  *****
\\
\hline
   0.00
 &     0.3183
 &     0.2927
 &     0.2674(2) 
 &     0.2420(2) 
 &     0.2162(4) 
 &     0.1895(7) 
 &      0.161(2) 
 &     *****
 &     *****
  \\
   0.01
 &     0.3183
 &     0.2930(1) 
 &     0.2678(2) 
 &     0.2425(3) 
 &     0.2167(4) 
 &     0.1903(4) 
 &     0.1630(5) 
 &       0.80(2) 
 &       0.92(5) 
  \\
   0.02
 &     0.3183
 &     0.2932
 &     0.2681(1) 
 &     0.2428(1) 
 &     0.2171(1) 
 &     0.1907(2) 
 &     0.1633(2) 
 &      0.712(6) 
 &      0.789(8) 
  \\
   0.03
 &     0.3183
 &     0.2934
 &     0.2684
 &     0.2432
 &     0.2174
 &     0.1910
 &     0.1637
 &      0.666(2) 
 &      0.728(3) 
  \\
   0.04
 &     0.3183
 &     0.2937(2) 
 &     0.2690(3) 
 &     0.2438(5) 
 &     0.2181(7) 
 &     0.1917(9) 
 &      0.164(1) 
 &       0.63(1) 
 &       0.69(1) 
  \\
   0.05
 &     0.3183
 &     0.2940
 &     0.2693
 &     0.2442
 &     0.2186
 &     0.1925
 &     0.1658(2) 
 &      0.612(2) 
 &      0.660(5) 
  \\
\end{tabular}
\end{ruledtabular}
\end{table*}

\end{document}